\documentclass[12pt, preprint]{aastex}
\usepackage{epsf}
\usepackage{times}
\newcommand{\etal}{{et al}\/.}

\begin{document}
\slugcomment{Draft of \today}
\shorttitle{Multiple hotspots in radio galaxies}
\shortauthors{M.J.\ Hardcastle \etal}
\title{A {\it Chandra} study of particle acceleration in the multiple
  hotspots of nearby radio galaxies}
\author{M.J.\ Hardcastle and J.H. Croston}
\affil{School of Physics, Astronomy \& Mathematics, University of
  Hertfordshire, College Lane, Hatfield AL10 9AB, UK}
\and
\author{R.P. Kraft}
\affil{Harvard-Smithsonian Center for Astrophysics, 60 Garden Street, Cambridge, MA~02138, USA}
\begin{abstract}
\noindent We present {\it Chandra} observations of a small sample of
nearby classical double radio galaxies which have more than one radio
hotspot in at least one of their lobes. The X-ray emission from the
hotspots of these comparatively low-power objects is expected to be
synchrotron in origin, and therefore to provide information about the
locations of high-energy particle acceleration. In some models of the
relationship between the jet and hotspot the hotspots that are not the
current jet termination point should be detached from the energy
supply from the active nucleus and therefore not capable of
accelerating particles to high energies. We find that in fact some
secondary hotspots are X-ray sources, and thus probably locations for
high-energy particle acceleration after the initial jet termination
shock. In detail, though, we show that the spatial structures seen in
X-ray are not consistent with na\"\i ve expectations from a simple
shock model: the current locations of the acceleration of the
highest-energy observable particles in powerful radio galaxies need
not be coincident with the peaks of radio or even optical emission.
\end{abstract}

\section{Introduction}
\label{intro}

In the standard model for powerful extragalactic double radio sources
(also known as classical double or FRII [\citealt{fr74}]
sources), hotspots, the bright compact regions at the ends of the
source, are the visible manifestation of a strong shock as the
relativistic beam of energetic particles is suddenly
decelerated by interaction with the slow-moving or stationary plasma
within the radio lobes \citep[e.g.,][]{br74}. The particle
acceleration at these shocks determines the energy distribution of the
electrons (and, possibly, protons) that go on to form the large-scale
lobes and expand into the external medium, and so an understanding of
how and where it happens is essential to an understanding of the
dynamics and environmental impact of radio sources; in addition, the
strong shocks in FRIIs are often invoked as a possible region of
acceleration for the high-energy cosmic ray population, so that it is
important to understand where (and if) high-energy particles are
accelerated in these systems.

The best evidence for this shock model comes from the radio through
optical spectra of hotspots, which have been shown
\citep[e.g.,][]{mrhy89} to be commonly consistent with the predictions
of a simple model for shock particle acceleration and downstream
losses \citep{hm87}. However, the idea that the hotspots always trace
the shock at the jet termination is challenged by the observation that
the lobes of radio galaxies and quasars very frequently have more than
one compact bright radio feature that meets whatever definition of a
hotspot is in use \citep[e.g.,][]{lbdh97}. Where these appear in the
jet or embedded deep in the lobe they are usually interpreted as `jet
knots' --- the assumption encoded in this terminology is that they are
telling us about internal dissipation in the jet rather than
disruption and that they are not relevant to the particle acceleration
history of the source. But in many systems there are multiple hotspots
at the far end of the lobe, and in these the configuration of the
hotspots relative to the jet flow often suggests that more than one is
associated with the beam termination. Models to explain this
observation include those in which the beam end-point moves about from
place to place in the lobe (the `dentist's drill' model of
\citealt{s82}) or in which material flows out from the initial impact
point of the beam to impact elsewhere on the lobe edge (the
`splatter-spot' model of \citealt{wg85} or the jet-deflection
model of \citealt{lb86}). All these models predict that
one of the hotspots, the one associated with the first or current
termination of the jet, should be more compact than the other or
others; it is in fact observed in the radio that where jets are
explicitly seen to terminate, they always do so in the most compact,
`primary' hotspot \citep[e.g.,][]{lbdh97,hapr97}.
But it is very difficult to distinguish between different models of
multiple hot spot formation using radio data alone. The secondary
(less compact) hotspots exhibit a wide variety of structures and
relationships to the primary hotspots, and, while in some cases radio
structure seems to favor one model rather than another, the nature of
the secondaries is never unambiguously constrained by single-frequency
radio data. One area in which different models {\it do} make different
predictions is that of the high-energy particle acceleration in the
secondary (less compact) hotspots. If these are relics left behind by
the motion of the jet, as in the `dentist's drill' model, then in
general we expect shock-driven particle acceleration to have ceased
(though the secondary hotspot may continue to be fed for some time if
the jet is disconnected some way upstream: \citealt{cgs91}).
Synchrotron losses will then deplete the high-energy electrons in the
secondary hotspot. If secondary hotspots continue to be fed by outflow
from the primary hotspot, then there is still an energy supply and
particle acceleration will continue to operate.

High-resolution X-ray observations of synchrotron radiation with {\it
Chandra} have provided key insights into the particle acceleration in
the jets of low-power, FRI-type radio galaxies, allowing us to locate
the sites of particle acceleration and relate them to the dynamics of
individual jets \citep[e.g.][]{hwkf03}. X-ray observations are
vital in these cases because the synchrotron loss timescale for
X-ray-emitting electrons is tens of years, assuming field strengths
close to the equipartition values: thus, unlike
radio and even optical observations of synchrotron radiation, X-ray
synchrotron detections tell us where particle acceleration is
happening {\it now}, rather than where it has happened in the past. To
date, however, it has been difficult to use X-ray observations to
study particle acceleration in the hotspots of the more powerful FRII
radio sources, because of the importance in hotspots of a second
emission process, inverse-Compton emission. This process, particularly
important in bright, compact hotspots, traces the low-energy electrons
rather than the high-energy ones, and has been the subject of much
work with {\it Chandra} because of its potential to measure physical
conditions (magnetic field strengths and energy densities) in hotspots
\citep*[see, e.g.,][]{hnpb00,hbw01,bbcp01}. Some hotspots have been known since the {\it ROSAT} epoch to be
best described by a synchrotron rather than inverse-Compton model
\citep*[e.g.][]{hll98} but until recently it has not been clear
what controls the relative dominance of the two processes. Based on
new and archival observations of a large sample of FRIIs, we recently
showed \citep{hhwb04} that the controlling parameter is
related to the overall {\it luminosity} of the hotspot:
high-luminosity hotspots never show X-ray synchrotron emission, while
low-luminosity hotspots often do. We argued that this is due to the
higher magnetic field strengths and photon energy densities found in
the more luminous hotspots: these increase the energy loss rate for
high-energy electrons and prevent efficient particle acceleration to
the energies needed for X-ray synchrotron radiation. In contrast,
low-luminosity hotspots can readily accelerate particles to X-ray
emitting energies, and the expected inverse-Compton emission is
negligible, so that synchrotron radiation is dominant in these
systems. It is important to note that this picture, while
qualitatively plausible, relies on details of the microphysics, such
as the magnetic field configuration and electron diffusion coefficient
in the acceleration region \citep[see, e.g.,][]{bmpv03} and so it
cannot yet be shown quantitatively to be correct.

If this model is accepted, though, X-ray synchrotron emission can be
used to probe high-energy particle acceleration in low-luminosity FRII
radio galaxies. An example of this is provided by observations of the
low-power FRII 3C\,403 \citep{khwm05}. These observations were
important for two reasons. Firstly, as 3C\,403 is a narrow-line radio
galaxy, any relativistic beaming effects (of the kind thought to be
important in some core-dominated quasars) must be minimal if unified
models are correct, as 3C\,403 should lie close to the plane of the
sky: thus, models for anomalous X-ray emission involving beaming, like
those of \citet{gk03}, need not be considered. Secondly, 3C\,403's E
lobe is a multiple-hotspot system, and the observations show a clear
difference between features of the jet and the primary hotspot, on the
one hand, and the secondary hotspot (much brighter in the radio), on
the other: we found that the upper limit on the X-ray emission from
the secondary was an order of magnitude below what would have been
detected if its X-ray to radio ratio had been the same as that in the
primary. This strongly suggests that, in this source at least, the
secondary hotspot is unable to accelerate particles to the highest
observable energies.

The results of our work on 3C\,403 motivated us to carry out further
observations of nearby sources with multiple hotspots, and to examine
data available in the {\it Chandra} archive, with the aim of seeing
whether, and in what circumstances, the different hotspot components
can give rise to high-energy particle acceleration, and so
constraining the nature of multiple hotspots. In this paper we report
on our results.

Throughout the paper we use a cosmology with $H_0 = 70$ km s$^{-1}$
Mpc$^{-1}$, $\Omega_{\rm m} = 0.3$ and $\Omega_\Lambda = 0.7$. The
spectral index $\alpha$ is defined in the sense that flux density
$\propto \nu^{-\alpha}$: the relationship between the spectral index
$\alpha$ and the photon index $\Gamma$ is thus $\Gamma = \alpha + 1$.

\section{Observations}

\subsection{Sample}

We selected our targets from the sample of Leahy \etal\ (1997:
hereafter L97). This sample consists of FRII radio galaxies with
$z<0.15$ taken from the 3CR sample (\citealt{sdma85}: see L97 for
details of the selection). We chose this parent sample because of its
low redshift (and therefore, in general, low luminosity, implying
negligible X-ray inverse-Compton radiation from the hotspots in the
picture of \citealt{hhwb04}) and because of the availability of
excellent radio data, with resolution matched to or exceeding that of
{\it Chandra}, for almost all members of the sample. From the L97
sample, we selected the sources with clear, well separated, bright
multiple hotspots as seen in the radio maps. Two of these, 3C\,390.3
and 3C\,403, had already been observed with {\it Chandra} (as
discussed in \citealt{hc05} and \citealt{khwm05}
respectively). We were awarded time for two more, 3C\,227 and 3C\,327,
giving us two broad-line and two narrow-line radio galaxies in total.
Basic properties of the sample objects are given in Table
\ref{sample}. We discuss the radio and X-ray data for these objects in
the following two sections.

\begin{table}
\caption{Properties of the sample of FRII sources}
\label{sample}
\begin{tabular}{lrlrrrr}
\hline
Name&$z$&Emission-line&$L_{178}$&Angular scale&Largest hotspot&$N_{\rm H}$\\
&&type&(W Hz$^{-1}$ sr$^{-1}$)&(kpc arcsec$^{-1}$)&separation (kpc)&(cm$^{-2}$)\\
\hline
3C\,227&0.0861&BLRG&$4.7 \times 10^{25}$&1.61&17&$2.08 \times 10^{20}$\\
3C\,327&0.1039&NLRG&$8.1 \times 10^{25}$&1.91&5&$6.49 \times 10^{20}$\\
3C\,390.3&0.0561&BLRG&$3.1 \times 10^{25}$&1.09&28&$3.68 \times 10^{20}$\\
3C\,403&0.059&NLRG&$1.9 \times 10^{25}$&1.14&4&$1.54\times 10^{21}$\\
\hline
\end{tabular}
\vskip 5pt
\begin{minipage}{15cm}
References for Galactic column densities used are as follows: 3C\,227,
\citet{jmdl85}; 3C\,327, interpolated from results of \citet{sgwb92};
3C\,390.3, \citet{mlle96}; 3C\,403, from \citet{khwm05}.
\end{minipage}
\end{table}

\subsection{Chandra observations}

Details of the {\it Chandra} observations of our targets are given in
Table \ref{chandra-obs}.

All {\it Chandra} observations were taken with the ACIS-S array with
the aimpoint, as standard, on the S3 chip. For 3C\,403 and 3C\,390.3
the nucleus was positioned near the aimpoint. However, for 3C\,227 and
3C\,327, our new targets, the (brighter) double hotspots were placed
near the aimpoint. This had the effect that the sources spanned
multiple {\it Chandra} chips. 3C\,227 and 3C\,327 were also observed
in Very Faint (VF) mode to reduce the background levels, while
3C\,390.3 and 3C\,403 were observed in Faint (F) mode. The
observations of 3C\,227 were split into two segments for scheduling
reasons.

\begin{table}
\caption{{\it Chandra} observations used in this paper}
\label{chandra-obs}
\begin{tabular}{lrlrrl}
\hline
Source&Obs. ID&Date&Obs. time (ks)&Filtered livetime (ks)&Obs. mode\\
\hline
3C\,390.3&830&2000 Apr 17&36.2&29.4&F\\
3C\,403&2968&2002 Dec 07&50.5&44.8&F\\
3C\,227&7265&2006 Jan 11&21.8&21.8&VF\\
       &6842&2006 Jan 15&31.2&31.2&VF\\
3C\,327&6841&2006 Apr 26&40.2&40.2&VF\\
\hline
\end{tabular}
\end{table}

We did not re-analyse the data for 3C\,403, as all the relevant
measurements had already been made \citep{khwm05}. For the other
three sources we re-processed the {\it Chandra} data with the latest
versions of {\sc ciao} and {\sc caldb} at the time of writing
(versions 3.3 and 3.2.3 respectively), following the standard {\sc
ciao} procedures. We removed the 0.5-pixel event position
randomization (since high spatial resolution is important to us) and
applied VF mode cleaning to the data for 3C\,227 and 3C\,327 to
improve the background. Intervals of high background count rate were
detected in the data for 3C\,390.3 using the {\it analyze\_ltcrv.sl}
script, and we removed these by time filtering the data.

The subsequent data analysis was carried out in {\sc ciao} and {\sc
xspec}. In what follows we present images in the 0.5--5.0 keV
passband, and carry out spectroscopy in the 0.4--7.0 keV band, unless
otherwise stated. The {\it specextract} script was used for spectral
extraction for extended sources, and the {\it psextract} script for
point sources. Spectral fits quoted for 3C\,227 are the results of
joint fitting to the data from both observations unless otherwise
stated. All errors quoted are $1\sigma$ for one interesting parameter.
The effects of the Galactic column density quoted in Table
\ref{sample} are included in all X-ray spectral fits.

\subsection{Radio data}

We have access to electronic versions of the maps of these sources
from \citet{bblp92} and \citet{lbdh97}. We made
radio maps using archival data at other frequencies where the existing
images were not adequate for our purposes. We also re-reduced the
8-GHz A-configuration data used by \citeauthor{bblp92} for imaging of
3C\,227, as the high-resolution maps available to us were sub-images
that did not show the core. Details of the radio data
we used are given in Table \ref{vlaobs}. All of the radio data was
reduced in {\sc aips} in the standard manner: individual datasets were
first self-calibrated and then cross-calibrated and merged with
appropriate weights to give a final multi-source dataset that was used
for imaging. Where maps at more than one resolution were required, we
applied appropriate tapering to the $uv$ plane in the imaging process.
Except where otherwise stated, maps used in the figures were made for
this paper.

\begin{table}
\caption{VLA data reduced for this paper}
\label{vlaobs}
\begin{tabular}{lrlllr}
\hline
Source&Frequency&VLA obs. id.&Configuration&Obs. date&Time on source\\
&(GHz)&&&&(h)\\
\hline
3C\,227&8.3&AB534&A&1990 May 25&3.0\\
&1.5&AS659&A&1999 Jul 10&1.1\\
&1.5&AS677&B&2000 Jan 29&1.5\\
&1.5&AZ28&C&1985 Sep 16&0.4\\[4pt]
3C\,327&1.5&AB376&A&1986 Mar 11&0.2\\
&1.5&AB376&B&1986 Aug 18&0.2\\
&1.5&AP77&C&1984 Apr 23&0.4\\[4pt]
3C\,390.3&5.0&VP88G&B&1989 Apr 17&10.4\\
&5.0&AS542&C&1994 Nov 27&0.9\\
\hline
\end{tabular}
\end{table}

As detailed comparison between the radio and X-ray spatial structures
of the hotspots is important to us, we aligned the radio and X-ray
frames for all the sources by shifting the X-ray nuclear position to
match the best available radio position. Where there were significant
offsets between the radio core positions at different frequencies, we
shifted the low-frequency positions to match the high-frequency ones.
We expect the relative astrometry of the radio and X-ray frames to be
limited by the accuracy of X-ray centroiding, but in all cases it
should be better than $\sim 0.1''$. 3C\,390.3 is a special case, as
extreme pileup has removed all counts from the center of the nuclear
X-ray emission: we still believe that we have been able to determine
the X-ray position to the required level of accuracy, and in this case
the default astrometry of the {\it Chandra} data appears to be correct.

For radio data, resolutions are quoted as the major $\times$ minor
axis (FWHM) of the elliptical restoring Gaussian: where only one
dimension is quoted the restoring beam used was constrained to be
circular.

\section{Results}

\subsection{Non-hotspot X-rays in 3C\,227 and 3C\,327}

In this subsection we briefly comment on the features of the new {\it
Chandra} observations of 3C\,227 and 3C\,327 unrelated to the
hotspots. The {\it Chandra} observations of the other sample sources
have been discussed elsewhere \citep[see][]{khwm05,hhwb04,hc05,ewhk06}.

\begin{figure}
\epsfxsize 16cm
\epsfbox{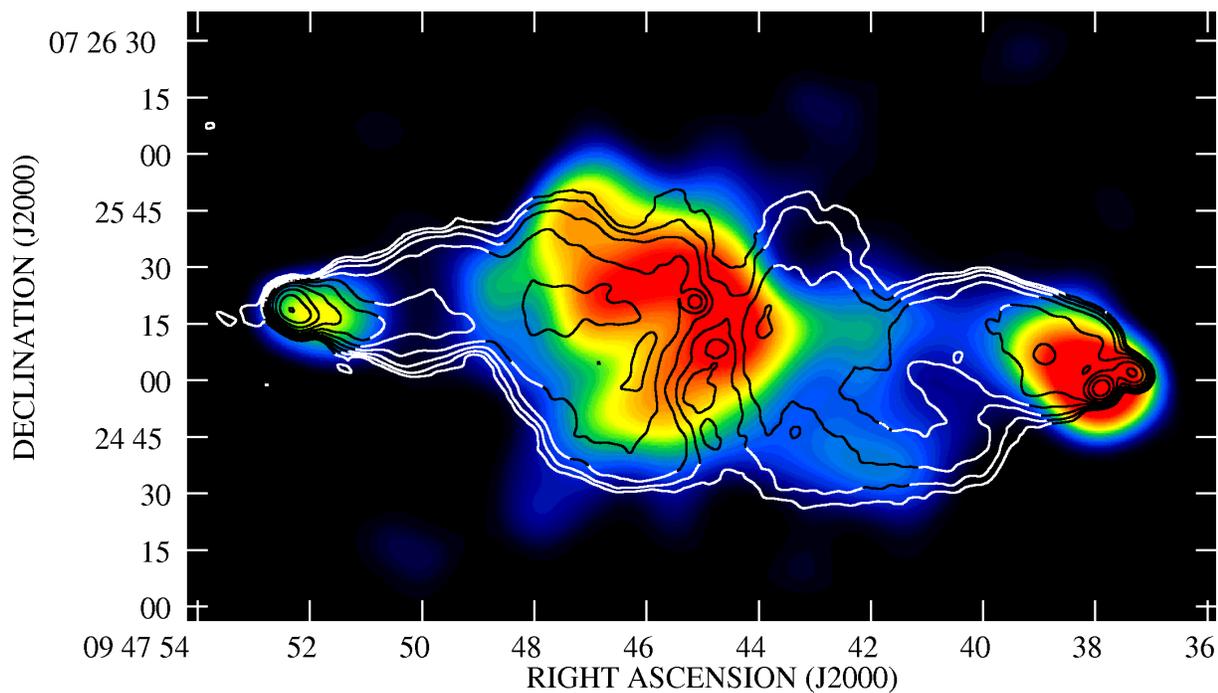}
\caption{Large-scale X-ray emission from 3C\,227. Background point
  sources and the central bright core (but not the hotspots) have been
  masked out and the resulting exposure-corrected image in the 0.5-5.0 keV passband
  smoothed with a Gaussian with FWHM $18\farcs5$. Overlaid are
  contours from our $4''$-resolution radio map at 1.5 GHz at $1 \times
  (1,2,4\dots)$ mJy beam$^{-1}$.}
\label{227-l}
\end{figure}

\begin{figure}
\epsfxsize 16cm
\epsfbox{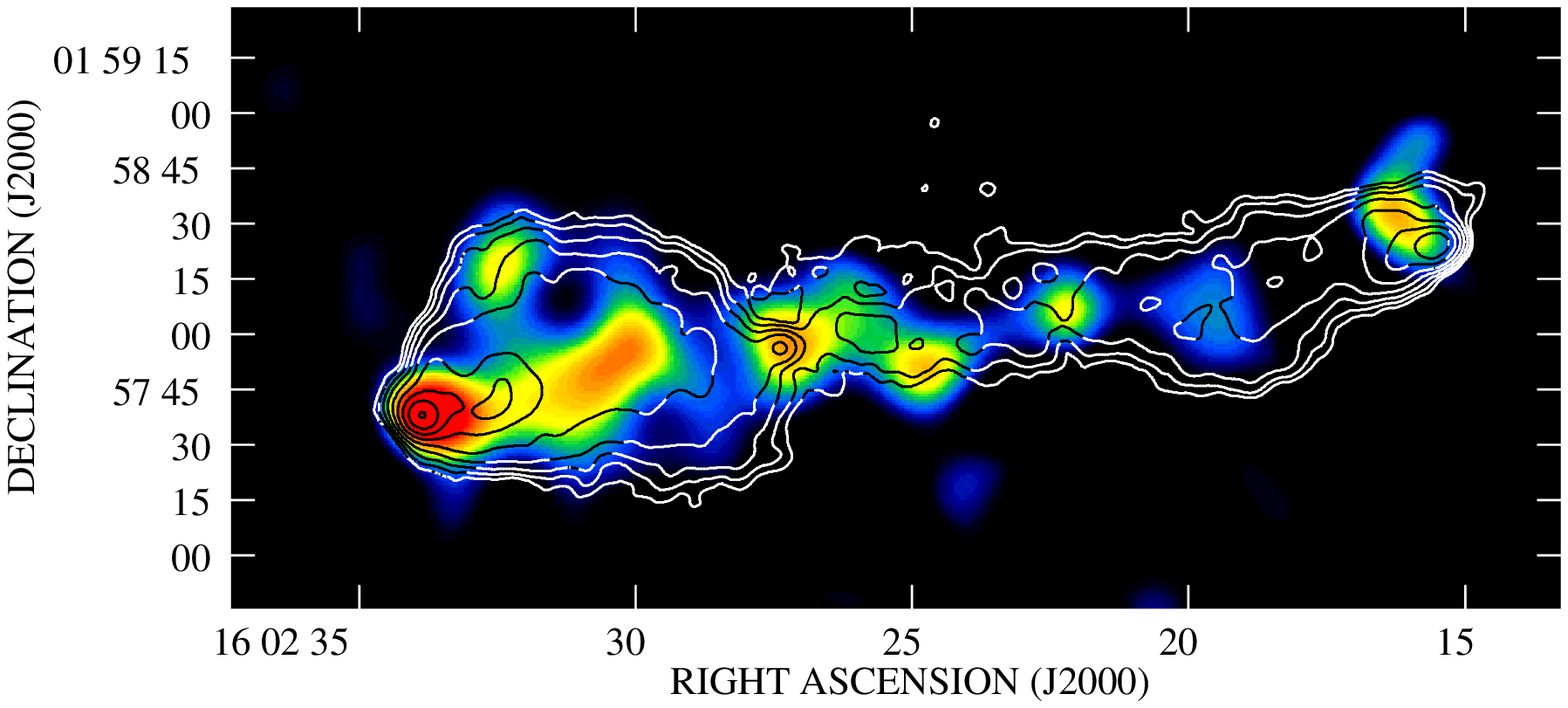}
\caption{Large-scale X-ray emission from 3C\,327. Background point
  sources and the central bright core (but not the hotspots) have been
  masked out and the resulting exposure-corrected image in the 0.5-5.0 keV passband
  smoothed with a Gaussian with FWHM $18\farcs5$. Overlaid are
  contours from our $6''$-resolution radio map at 1.5 GHz at $2 \times
  (1,2,4\dots)$ mJy beam$^{-1}$.}
\label{327-l}
\end{figure}

Images showing the large-scale X-ray emission from the two objects are
shown in Figs \ref{227-l} and \ref{327-l}.

\subsubsection{Cores}

The raw {\it Chandra} count rate from the core of the BLRG 3C\,227 is
roughly 0.14 s$^{-1}$: with a frame time of 3.1 s, this means that we
might expect it to be affected by pileup at a significant
level\footnote{http://cxc.harvard.edu/proposer/POG/html/ACIS.html}.
Fitting to the total spectrum of the nucleus within 15 {\it Chandra}
pixels (7 arcsec), using a concentric adjacent background annulus, we
found a good fit ($\chi^2 = 228$ for 213 d.o.f.) to a model consisting
of two power laws, one with Galactic absorption only and one with
additional intrinsic absorption at the redshift of the source. The
power-law component without intrinsic absorption had a steep
best-fitting photon index ($\Gamma = 3.4_{-0.2}^{+0.6}$) while the
absorbed component, with $N_{\rm H,int} = (1.4 \pm 0.1) \times
10^{22}$ cm$^{-2}$, had a very flat photon index, $\Gamma = 0.73 \pm
0.06$. The data are almost equally well fitted ($\chi^2 = 231/214$)
with an ionized absorber model ({\sc absori}) but this has clear
residuals at soft energies, suggesting that in any event some of the
soft X-ray emission must come from a second nuclear X-ray component,
or from soft thermal X-ray emission close to the nucleus. This is
unsurprising as we know that in general radio-loud AGN have a soft
X-ray component which is probably related to the jet
\citep*[e.g.,][]{hec06}. The significant intrinsic absorption seen
here, although unusual for a broad-line object, is consistent with
what is seen in some other BLRG or quasars (e.g. 3C\,109,
\citealt{afii97}, \citealt{hec06}; 3C\,351, \citealt{nfpe99},
\citealt{hbch02}). From optical spectropolarimetry \citet{cotg99}
estimate that the optical broad lines and continuum may be obscured by
1-2 magnitudes in the $V$ band. Although our column density would
imply substantially higher extinction, $A_V \sim 6$ for Galactic
gas/dust ratios, the optical results, including the detection of
polarized broad emission lines, give us some additional reason for
believing that there might be intrinsic X-ray absorption in this
source.

The very flat photon index of the absorbed component is likely to be
partly the result of pileup, and to investigate this we also extracted
a spectrum for an annular region between 2 and 15 {\it Chandra}
pixels, with the same background region. This excludes the region
where pileup is likely to be significant. We applied energy-dependent
corrections to the ARF using the algorithm of the {\sc arfcorr}
software\footnote{http://agyo.rikkyo.ac.jp/$\sim$tsujimot/arfcorr.html}
implemented using the {\sc funtools} package. Fitting the same double
power-law spectrum, we find similar parameters --- the slightly larger
intrinsic absorbing column ($1.7_{-0.3}^{+0.5} \times 10^{22}$
cm$^{-2}$) is consistent within the errors --- and only a slightly
steeper photon index for the absorbed component, $0.91 \pm 0.18$.
However, the 99\% confidence upper limit on the photon index in this
extraction region (which obviously contains only a small fraction of
the total counts) is 2.4, so more typical photon indices are certainly
not excluded by the non-piled up data.
\label{227-core}

\citet{cf95}, who discuss observations of 3C\,227 with
the {\it ROSAT} PSPC , found that the nuclear spectrum was well fitted
by a power-law model with $\Gamma \sim 1.5$. With the limited signal
to noise of their {\it ROSAT} data and the restricted energy range of
the PSPC it is not clear that they could have identified the
absorption features seen in our spectrum.

For 3C\,327, which has a much fainter X-ray nucleus, we extracted a spectrum
from a source circle of radius 4 {\it Chandra} pixels (2 arcsec) with
an adjacent concentric background annulus. The best-fitting spectrum
($\chi^2 = 35$ for 21 d.o.f.) again requires multiple components: at
soft energies the source is dominated by an unabsorbed power law with
$\Gamma = 3.07 \pm 0.15$, while the residuals at high energy require
an additional heavily absorbed power law, with the poorly constrained
$\Gamma$ fixed at 1.7 \citep[following][]{hec06} and $N_{\rm
H,int} = (6_{-2}^{+5}) \times 10^{22}$ cm$^{-2}$, together with a strong
line-like feature modeled as a Gaussian with $E = 6.42 \pm 0.05$ keV
and $\sigma = 0.19 \pm 0.08$ keV. Except for the 
steep power law index for the unabsorbed component, this is a typical spectrum
for a narrow-line radio galaxy: iron features around 6.4 keV are often
found in these objects (\citealt*{sem99}; \citealt{ewhk06}; \citealt{hec06}). 

\subsubsection{Lobes}

The lobes of 3C\,227 are both clearly detected in the {\it Chandra}
observations (Fig. \ref{227-l}). Extended X-ray emission from lobes of
radio sources is generally attributed to inverse-Compton scattering
from the microwave background radiation (CMB)
\citep{hbch02,ks05,chhb05}. We extracted a spectrum for the W lobe, as
the E lobe spans a chip boundary, using a polygonal extraction region
defined by the radio emission but excluding the central part of the
lobe where the nucleus and central extended X-ray emission might
contribute, as well as compact sources, including the hotspots. The
spectrum was well fitted ($\chi^2 = 10.6$ for 10 d.o.f.) with a
power-law model with $\Gamma = 1.77 \pm 0.26$ and 1-keV unabsorbed
flux density of $4.8 \pm 0.5$ nJy. We used the code of \citet*{hbw98}
to calculate the predicted inverse-Compton emission from the whole
lobe, modeling the lobe crudely as a uniform cylinder with length
$98''$ and radius $26''$, assuming equipartition, and using the same
electron spectral assumptions as \citet{chhb05}, with a low-energy
electron energy index (`injection index'), $p$, of 2. The
equipartition magnetic field strength is 0.58 nT; the predicted
emission corresponds to a 1-keV flux density of 1.4 nJy, i.e. about a
factor 3.5 below what is observed. Reducing the field strength to 0.29
nT is required to produce all the observed X-ray emission by the
inverse-Compton process.

For 3C\,327 there is a clear detection of the E lobe and a weaker but
still significant detection of the W lobe. Again, we extracted a
spectrum for the lobe that does not span a chip boundary, in this case
the E lobe, excluding point sources and the components near the
hotspots. This was well fitted ($\chi^2 = 6.9$ for 6 d.o.f.) with a
power-law model with $\Gamma = 1.55 \pm 0.30$ and unabsorbed 1-keV
flux density of $6.4 \pm 1.1$ nJy. The inverse-Compton prediction here
on the same assumptions (modeling the lobe as a cylinder with length
$90''$ and radius $27''$) is 1.5 nJy, so again the observations exceed
the equipartition prediction by a factor $\sim 4$. The equipartition
field here is 0.73 nT and the field required to produce the observed
X-rays by inverse-Compton processes is 0.34 nT.

Both objects are thus consistent with the trend seen in many other
sources for the observed inverse-Compton emission to lie somewhat
above the equipartition, $p=2$ prediction \citep{chhb05}. We
know that 3C\,227's lobes are likely to make a relatively small angle
to the line of sight, but this makes only a small difference to the
inverse-Compton prediction for $\theta = 45^\circ$.

%\subsubsection{Extended emission}
%
%3C\,227 has a well-known large extended emission-line region (Prieto
%et al 1993). Such regions in more powerful objects are often
%associated with extended X-ray emission (Crawford \etal\ ...).

\section{Hotspot observations}

In this section we discuss the X-ray, radio and (where possible)
optical properties of the hotspots in our target systems. Table
\ref{hstable} gives a summary of the properties discussed below for
ease of reference.

\begin{deluxetable}{llllrrrrlll}
\tablewidth{21cm}
\tabletypesize{\footnotesize}
\rotate
\tablecaption{Properties of the hot spot components discussed in the
  paper}
\label{hstable}
\tablehead{\colhead{Source}&\colhead{Lobe}&\colhead{X-ray}&\colhead{Radio}&\colhead{Offset}&\colhead{Counts}&\colhead{Photon}&\colhead{1-keV
  flux}&\colhead{Optical?}&\colhead{Type}&\colhead{Morphology}\\\colhead{name}&&\colhead{name}&\colhead{name}&&\colhead{(0.5-5.0
  keV)}&\colhead{index}&\colhead{density (nJy)}}

\startdata
3C\,227&W&--&P1/2&$0\farcs8 \pm 0\farcs1$&$84\pm 9$&$1.63 \pm 0.22$&$1.5 \pm
0.2$&Y&Primary&Resolved, flattened\\
&&--&P3&$2\farcs3 \pm 0\farcs2$&$19 \pm 5$&1.63&$0.3 \pm 0.1$&Y&Secondary&Resolved\\
&&--&P4&$0\farcs15
\pm 0\farcs1$&$22 \pm 5$&1.63&$0.4 \pm 0.1$&N&Jet knot?&Slightly resolved\\
&E&--&F1&$0\farcs5 \pm 0\farcs5$&$10\pm 3$&1.63&$0.3 \pm 0.1$&Y&Primary&Compact + diffuse\\[4pt]
3C\,327&E&SX1&S1?&$2\farcs4 \pm 0\farcs1$&$12 \pm 3$&1.7&$0.27 \pm 0.07$&?&Primary&Compact\\
&&SX2&S2?&$0\farcs8 \pm 0\farcs1$&$12 \pm 3$&1.7&$0.27 \pm 0.07$&?&Secondary&Compact\\[4pt]
3C\,390.3&N&NX1&B&$0\farcs4 \pm 0\farcs1$&$164 \pm 13$&$1.95 \pm 0.15$&$4.2 \pm 0.3$&Y&Primary&2 sub-components\\
&&NX2&--&--&$72 \pm 9$&$2.13 \pm 0.24$&$1.9 \pm 0.2$&Y&?&Point-like\\
&&--&F&--&$<8$&1.95&$<0.2$&N&Secondary&No X-ray detection\\
&S&SX1&G$'$&$1\farcs0 \pm 0\farcs2$&$14 \pm 4$&1.4&$0.4 \pm 0.1$&N?&Primary?&Compact\\
&S&Diffuse&G,E,D?&--&$96 \pm 11$&$1.4 \pm 0.2$&$2.7 \pm 0.3$&Y?&Secondary&Diffuse\\[4pt]
3C\,403&E&--&F1&--&$34 \pm 6$&$1.75_{-0.3}^{+0.4}$&$0.9 \pm 0.2$&Y&Primary&Resolved\\
&&--&F1b&--&$15 \pm 4$&2.0&$0.5\pm 0.1$&N&Jet knot?&Compact\\
&&--&F2&--&$<4$&2.0&$<0.13$&N&Secondary&No X-ray detection\\
&&--&F6&--&$83 \pm 9$&$1.7_{-0.2}^{+0.3}$&$2.3 \pm 0.2$&Y&Jet knot&Elongated\\
\enddata
\tablecomments{Column 3 gives the name we have associated with the X-ray
  feature. Where none is given no specific name has
  been assigned in the text. Column 5 gives the offset between the
  peak of the radio counterpart named in column 4 and the centroid of
  the X-ray emission: it is blank if no radio counterpart is known.
  The photon index in column 7 is derived from the data if an error is
  quoted and is the value assumed in the text otherwise. Column 8 is
  the 1-keV flux density, determined from the spectral fit if
  possible, and otherwise from the count rate using the spectral
  assumption specified in column 7. Column 8 describes the nature of
  the radio hotspot, if known. Column 9 indicates whether there
  is an optical counterpart to the hotspot: Y = yes, N = no, ? = not
  known. Column 10 describes the X-ray structure: `compact' here
  indicates a weak source which is consistent with being $<0\farcs5$ in
  extent, and may be point-like. Measured parameters
  for 3C\,403 are taken from \citet{khwm05}: as no significant
  offsets are described in that paper no values are tabulated here.}
\end{deluxetable}

\subsection{3C\,227}

\begin{figure}
\epsfxsize 14cm
\epsfbox{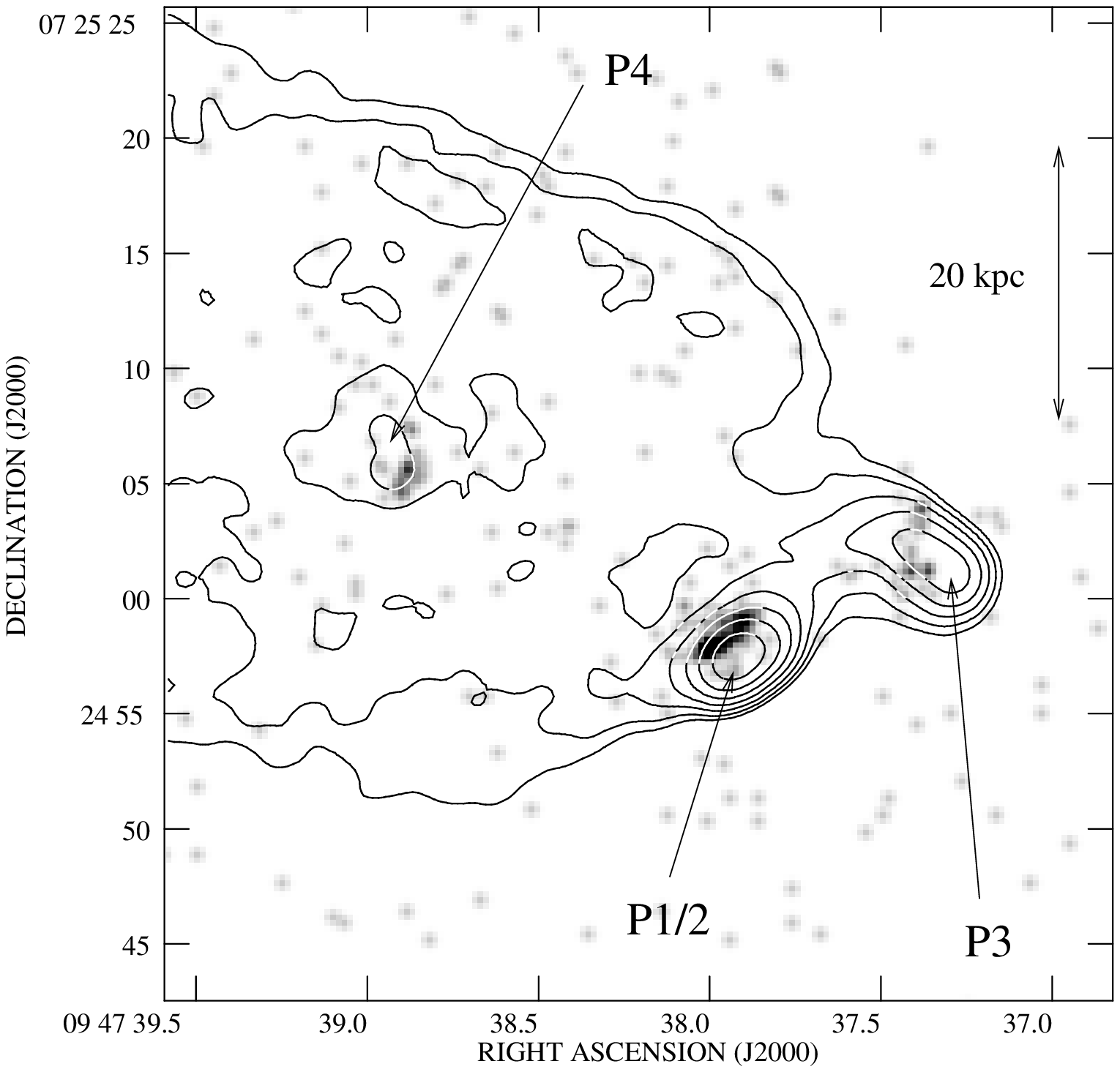}
\caption{The W hotspot of 3C\,227. The grayscale shows the merged
  0.5--5.0 keV {\it Chandra} data binned in $0\farcs246$ pixels and
  smoothed with a FWHM = $0\farcs5$ Gaussian. Contours are from our
  1.5-GHz radio map with $1\farcs5$ resolution, at $1 \times
  (1,2,4\dots)$ mJy beam$^{-1}$. Labels show the names of compact
  features in the notation of \citet{bblp92}.}
\label{227-hswl}
\end{figure}

\begin{figure}
\epsfxsize 14cm
\epsfbox{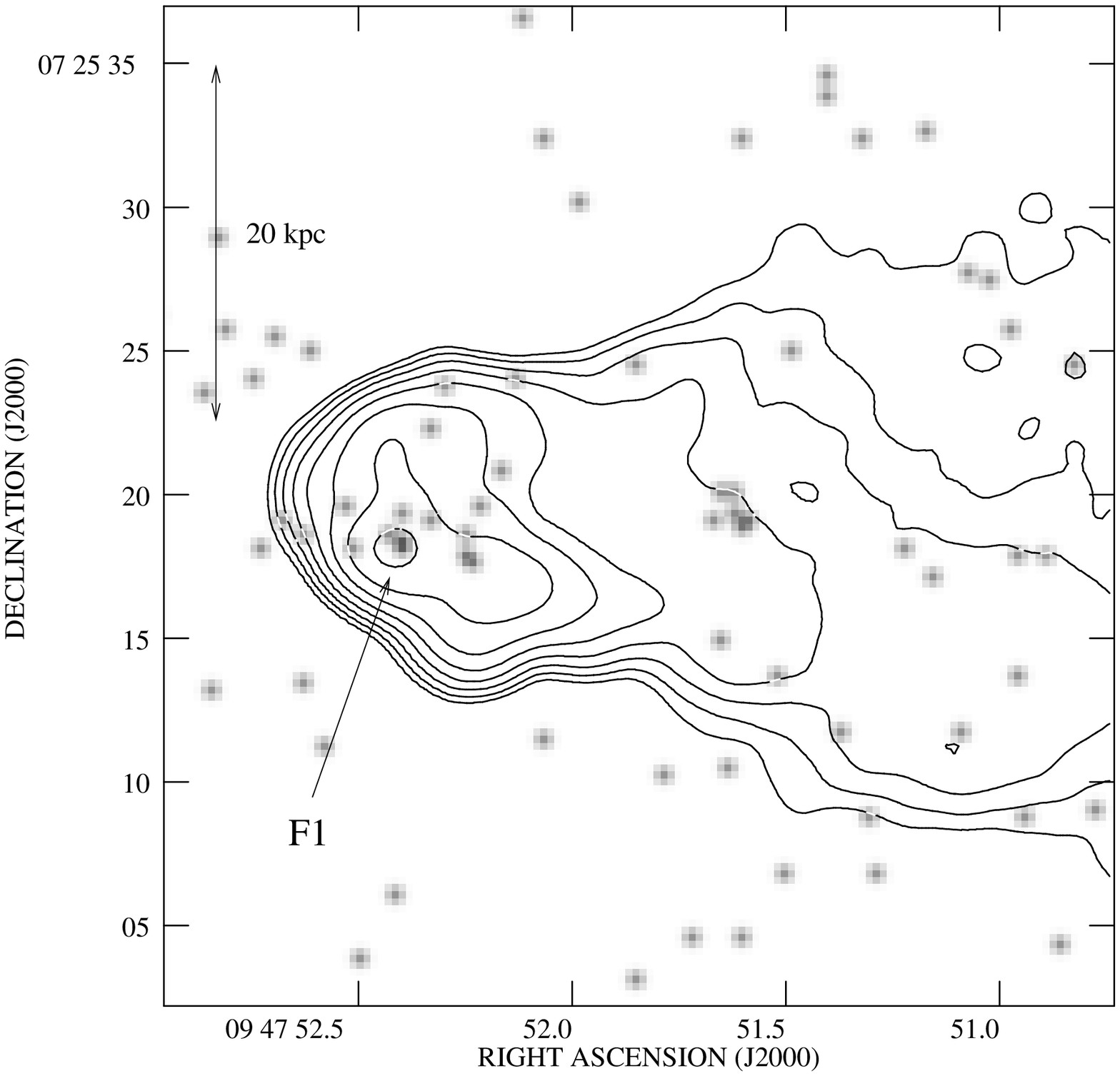}
\caption{The E hotspot of 3C\,227. The grayscale shows the merged
  0.5--5.0 keV {\it Chandra} data binned in $0\farcs246$ pixels and
  smoothed with a FWHM = $0\farcs5$ Gaussian. Contours are from our
  1.5-GHz radio map with $1\farcs5$ resolution, at $1 \times
  (1,2,4\dots)$ mJy beam$^{-1}$. Labels show the names of compact
  features in the notation of \citet{bblp92}.}
\label{227-hsel}
\end{figure}

\begin{figure}
\epsfxsize 14cm
\epsfbox{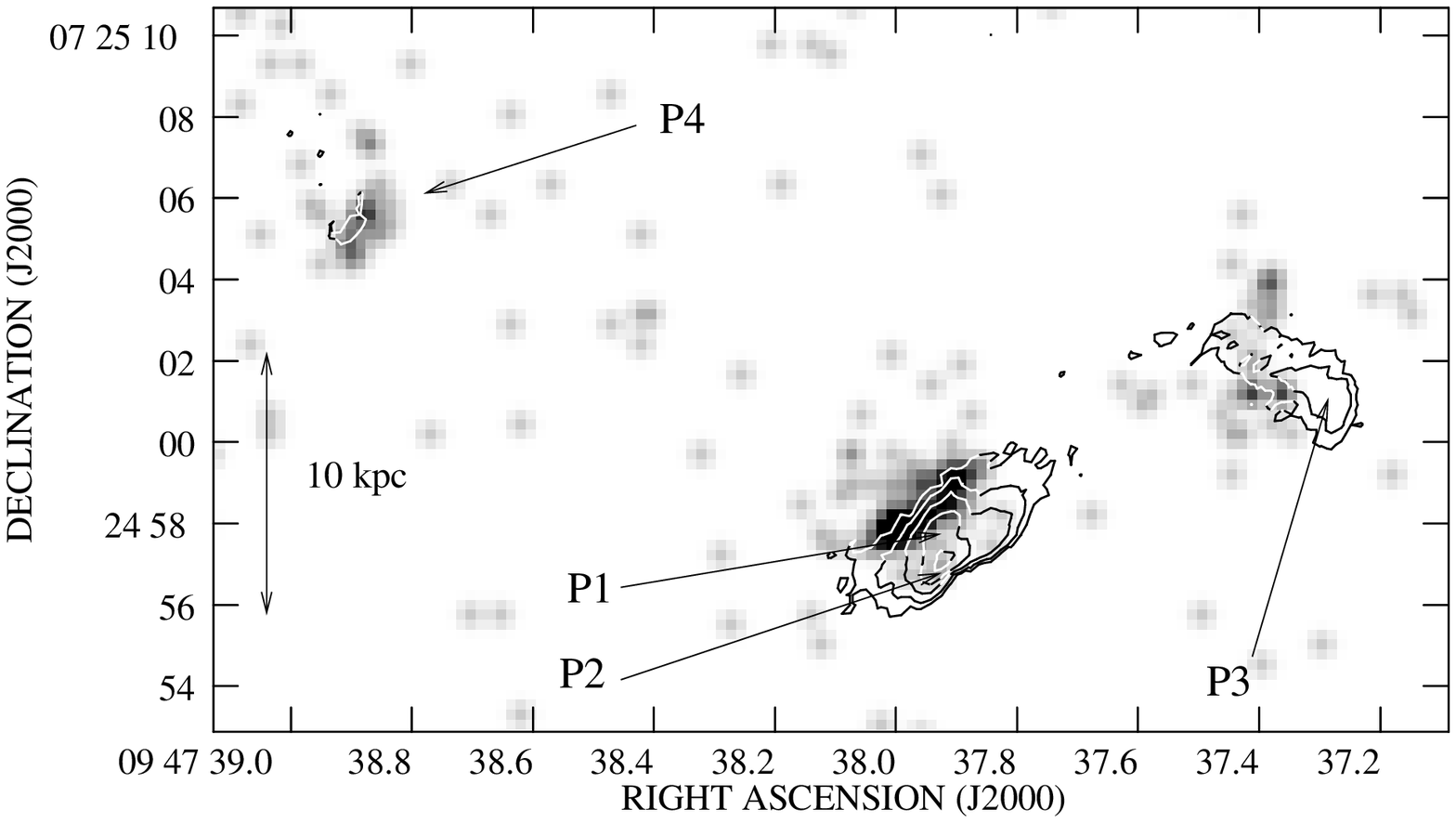}
\caption{The W hotspot of 3C\,227. The grayscale shows the X-ray data
  as in Fig.\ \ref{227-hswl}. Contours are from our 8.3-GHz radio map
  with $0\farcs37 \times 0\farcs22$ resolution, at $0.1 \times
  (1,2,4\dots)$ mJy beam$^{-1}$. Labels show the names of compact
  features in the notation of \citet{bblp92}.}
\label{227-hswx}
\end{figure}

Optical counterparts of 3C\,227's hotspots have recently been
discovered using ground-based observations \citep*{bmpv03,mpb03}. The
two components of the double W hotspot and the more compact E hotspot
all emit in the near-infrared. In the X-ray, similarly, emission is
detected from both hotspot regions, as shown in Figs \ref{227-hswl}
and \ref{227-hsel}. In the W hotspot X-ray emission is clearly
detected from the three compact components P1/2, P3 and P4, while
there is weak emission not just from the compact E component F1 but
also from the region around it. In the standard multiple hotspot
picture P4 would probably be designated a jet knot (cf knot F6 of
3C\,403, below), P1/2 would be the primary hotspot and P3 a secondary
hotspot. Thus 3C\,227 is an example of a system where the secondary
hotspot clearly is an X-ray source. What is most striking in the W
hotspot region is the offsets between the peak of the radio and the
peak of the X-ray in the X-ray detected components. This is most
clearly seen in Fig.\ \ref{227-hswx}, where the X-ray emission is
compared with the high-resolution radio data used by \citep{bblp92}.
At this resolution the brightest hotspot is resolved into two
components (P1 and P2), and both of these are associated with some
X-ray emission, but the brightest X-ray emission comes from a resolved
structure which is comparable in size to P2 (with a long axis of $3''$
or 5 kpc) and the X-ray centroid of this structure is displaced
$0\farcs8 \pm 0\farcs1$ (1.3 kpc) in the direction of the nucleus from
the peak of P1. The centroid of the X-ray emission from P3 is
displaced by $2\farcs3\pm 0\farcs2$ (3.6 kpc) from the peak of the
radio emisission in the direction of P1/2. Finally, the X-ray emission
from P4 appears slightly displaced ($0\farcs15 \pm 0\farcs11$) from
the peak of the radio in the direction of P1/2, but the statistics are
not good enough to claim a definite offset. In contrast, the centroid
of the emission from F1 and its surroundings is consistent with the
position of the radio peak within the large errors.

The X-ray counterpart of P1/2 contained enough counts ($84 \pm 9$ in
the 0.5--5.0 keV energy range) for us to extract a spectrum, which we
took from a circular region with radius 6 {\it Chandra} pixels (3
arcsec) with background from a concentric annular region. We find a
good fit ($\chi^2 = 0.5$ for 3 d.o.f.) with a power-law model with
$\Gamma = 1.63 \pm 0.22$ and 1-keV unabsorbed flux density $1.5 \pm
0.2$ nJy. The fainter hotspots P3 and P4 contain $19 \pm 5$ and $22
\pm 5$ net 0.5--5.0 keV counts respectively, which for the same
spectrum would correspond to flux densities of $0.3 \pm 0.1$ and $0.4
\pm 0.1$ nJy. The counterpart of the F1 hotspot, which lies on the S1
chip, contains $10\pm 3$ counts in a similarly sized extraction
region, which corresponds to $0.3\pm 0.1$ nJy if again we assume the
same spectrum as seen in P1/2. 

K.-H. Mack kindly provided us with electronic versions of
ground-based near-infrared (NIR) images (\citealt{mpb03}, and in
preparation). When shifted to our radio-based co-ordinate system by
alignment at the peak of the optical emission from the galaxy these
show that the optical emission from the P1/2 region extends over both
the radio and X-ray peaks. In detail, though, the brightest regions of
NIR emission are not coincident with either radio or X-ray peaks. The
NIR emission disappears in between the peak of the X-ray and the peak
of the radio emission at P2. Similarly, the peak of the NIR emission
coincident with P3 agrees neither with the radio nor the X-ray peak
positions, though it is roughly coincident with the X-ray centroid.
There is no NIR counterpart of the radio/X-ray knot P4.

\subsection{3C\,327}

\begin{figure}
\epsfxsize 14cm
\epsfbox{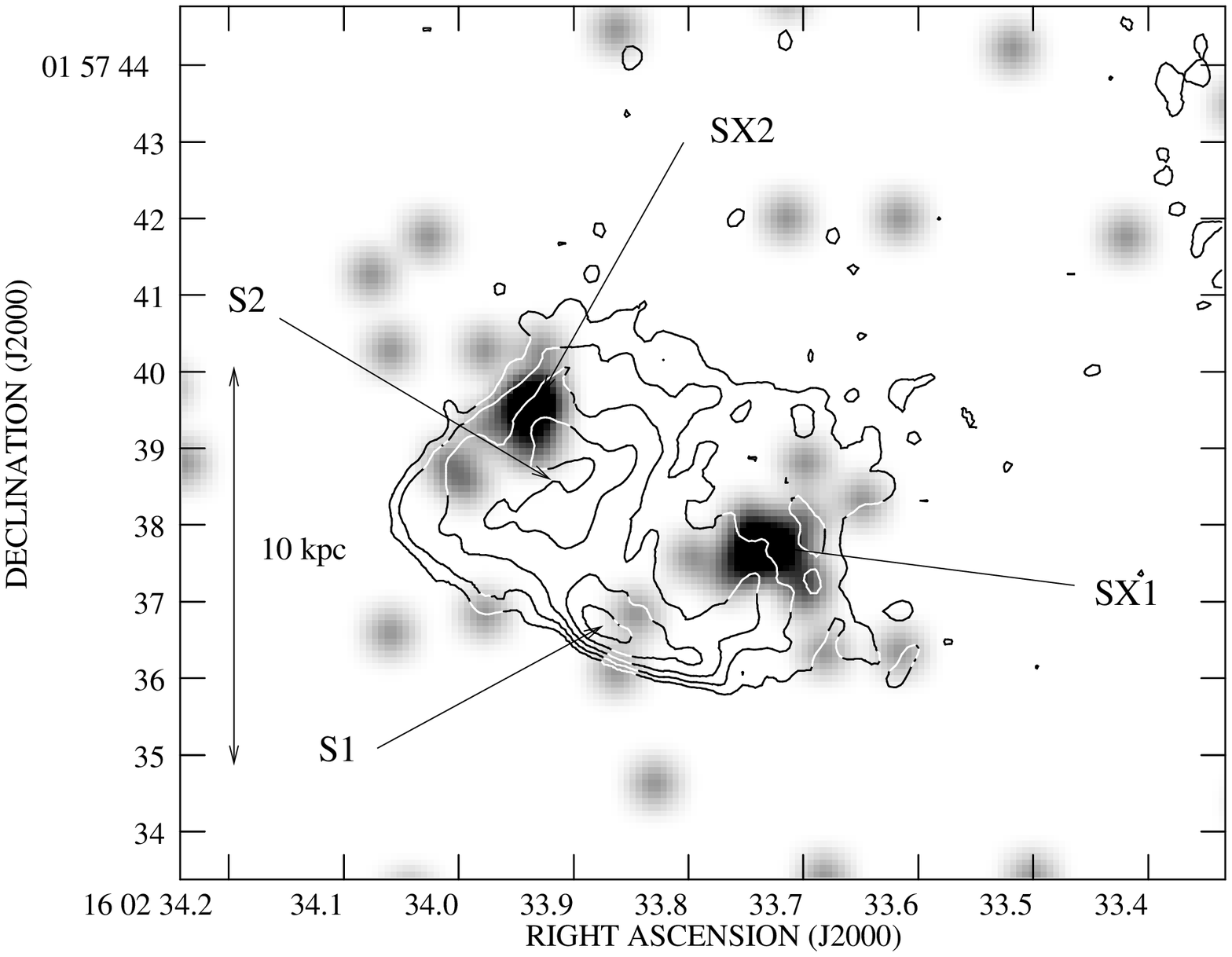}
\caption{The E hotspot of 3C\,327. The grayscale shows the
  0.5--5.0 keV {\it Chandra} data binned in $0\farcs246$ pixels and
  smoothed with a FWHM = $0\farcs5$ Gaussian. Contours are from the
  8.2-GHz radio map of \citet{lbdh97}
  with $0\farcs29$ resolution, at $0.2 \times
  (1,2,4\dots)$ mJy beam$^{-1}$. Labels show the names of compact
  radio features in the notation of \citet{lbdh97} and
  their (possibly) corresponding X-ray features.}
\label{327-hsex}
\end{figure}

3C\,327 is the only object in our sample to have no optical hotspot
detection: as pointed out by \citet{mpb03}, a nearby bright disk
galaxy makes it hard to detect potentially faint counterparts to the
bright double eastern radio hotspots. In the new {\it Chandra} data,
although there is nearby X-ray emission (Fig.\ \ref{327-l}) there is
no detection of a compact component corresponding to the W hotspot.
There is, however, a clear detection of compact components near the
double E hotspots (Fig. \ref{327-hsex}). Neither of the two X-ray
components detected bears a very obvious relation to the radio
structure. The component (denoted SX1) closest to the primary hotspot
(S1) is separated from it by $2\farcs4\pm0\farcs1$ (4.5 kpc) along the
line connecting S1 and the core, while the other X-ray component (SX2)
is at $0\farcs8 \pm 0\farcs1$ (1.5 kpc) from the nearest peak of the
radio emission in the S2 region, and considerably further from its
center, in the sense of being further away from the primary hotspot
than the peak of the radio emission. In DSS2 and 2MASS images SX2 is
close to the nearby disk galaxy, but not at its nucleus. We cannot
rule out the possibility that SX2 is associated with this galaxy
rather than with the 3C\,327 hotspot system. SX1 is further away from
the galaxy than SX2.

The two compact X-ray components each contain $12 \pm 3$ net 0.5--5.0
keV counts so that spectral fitting is not possible. If we assume a
power-law spectrum with $\Gamma = 1.7$, the observed net counts
correspond to flux densities of $0.27 \pm 0.07$ nJy in each component.
We estimated the hardness ratios of the two components (defining the
soft band as 0.5-2.0 keV and the hard band as 2.0-5.0 keV) and
compared them to the hardness ratio expected for a $\Gamma = 1.5$
power law with Galactic absorption (this model was chosen because the
spectral index is unlikely to be flatter than $\Gamma = 1.5$). We found
that SX2 is significantly harder (at the 99\% confidence level on a
binomial test) than would be expected for such a model, while SX1 is
not inconsistent with it. This suggests either that SX2 is not related
to the hotspot (e.g. that it is a background type 2 AGN) or that there
is some additional source of absorption, conceivably in the disk
galaxy, that affects SX2 but not SX1. We cannot distinguish between
these models with the available data.

\subsection{3C\,390.3}

\begin{figure}
\epsfxsize 14cm
\epsfbox{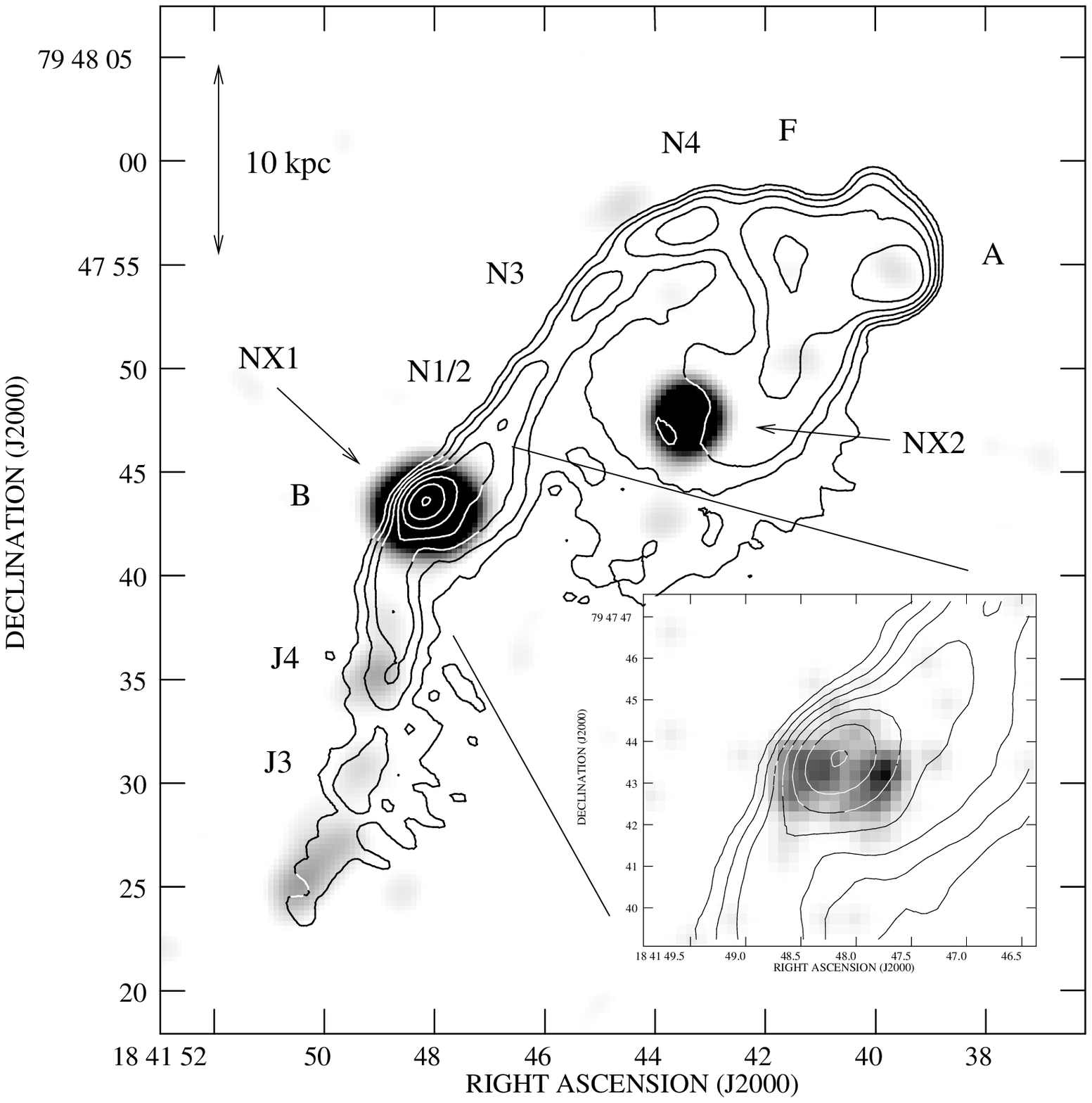}
\caption{The N hotspot of 3C\,390.3. The grayscale shows the 0.5--5.0 keV {\it Chandra} data binned in $0\farcs246$ pixels and
  smoothed with a FWHM = $2\farcs0$ Gaussian. Contours are from our
  5.0-GHz radio map, with $1\farcs0$ resolution, at $0.4 \times
  (1,2,4\dots)$ mJy beam$^{-1}$. Labels show the names of compact
  radio features in the notation of \citet{lp95} and the two
  compact X-ray features. The inset shows the same radio contours
  overlaid on the {\it Chandra} image of NX1 (the counterpart to knot
  B) smoothed with a FWHM = $0\farcs5$ Gaussian, showing that the
  X-ray source is resolved into two distinct knots.}
\label{390.3hsn}
\end{figure}

\begin{figure}
\epsfxsize 15cm
\epsfbox{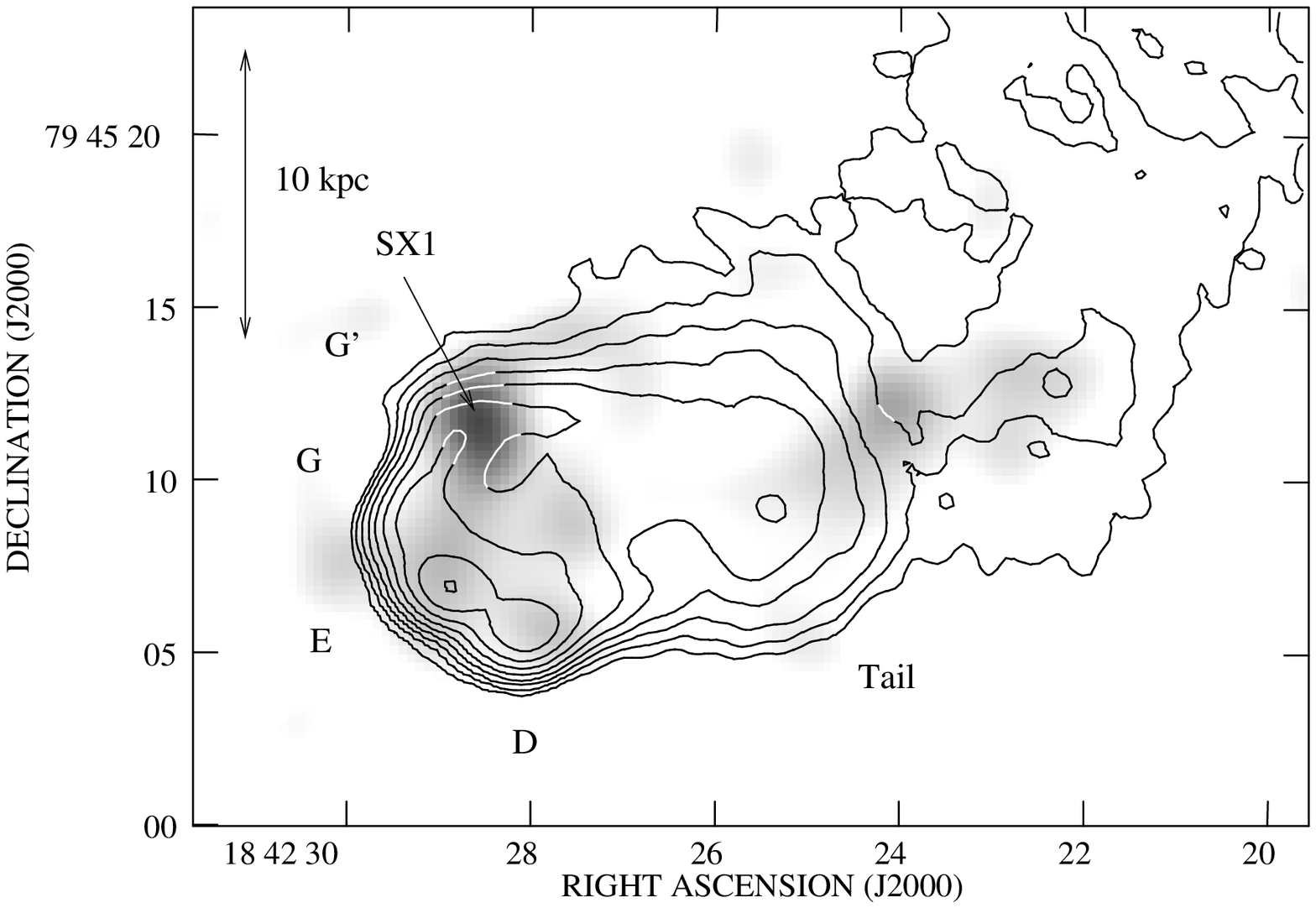}
\caption{The S hotspot of 3C\,390.3. The grayscale shows the 0.5--5.0 keV {\it Chandra} data binned in $0\farcs246$ pixels and
  smoothed with a FWHM = $2\farcs0$ Gaussian. Contours are from our
  5.0-GHz radio map, with $1\farcs0$ resolution, at $0.4 \times
  (1,2,4\dots)$ mJy beam$^{-1}$. Labels show the names of
  radio features in the notation of \citet{lp95} and of the
  one compact X-ray feature.}
\label{390.3hss}
\end{figure}

The northern compact hotspot of 3C\,390 is a well-known X-ray source
\citep{p97,hll98,hhwb04}. Diffuse X-ray emission from the large
hotspot region of the S lobe was discussed by \citet{hc05}. A detailed
examination of the {\it Chandra} data shows that several components of
the N hotspot complex are X-ray sources (Fig.\ \ref{390.3hsn}). The
radio jet, which is detectable throughout the N lobe \citep{lp95}
brightens strongly in the area shown by our image, and there is faint
but clear X-ray emission associated with this section of the jet. The
strongest X-ray `source' is the known X-ray counterpart of the primary
hotspot \citep{lp95}, knot B, here denoted NX1. The centroid of this
region is only slightly ($0\farcs4 \pm 0\farcs1$, 0.4 kpc) offset to
the S of the peak of the radio emission, but the resolution of Chandra
shows that the X-ray feature is actually resolved into two components,
separated by 1.5 arcsec (1.6 kpc) and placed symmetrically on either
side of the radio peak. This is a much smaller distance than that of
the nearby galaxy, possibly interacting with the jet, described by
\citet{hll98}, which is $5\farcs8$ to the NE. At higher radio
resolutions (e.g., in Figure 5 of \citealt{lp95}) knot B is resolved
into a linear structure which lies between the two X-ray peaks --
there is no sign of any double structure in the radio. Nor is there
any evidence in the optical image of \citet{pk97} or in the archival
{\it Spitzer} data for resolution of the hotspot in the E-W direction:
\citeauthor{pk97} show that the optical hotspot is in fact extended in
the same direction as the radio. Another bright X-ray source in the
hotspot region is the object we denote NX2. This has no radio
counterpart and is well separated from any compact structure in the
radio, so it seems most likely to be a chance superposition with a
background object, although it is detected in the optical and infrared
and its properties at these wavelengths are consistent with it being
similar to knot B (its infrared colors mean that it is certainly not
a background normal galaxy). Radial profile analysis shows that it is
consistent with being a point source. Finally, the secondary hotspot
(knots F and A) shows no significant X-ray emission, even though in
3C\,390.3 there is clear evidence (in the form of the continued
collimated outflow from knot B, features N1--4 in Fig.\
\ref{390.3hsn}) that this hotspot is connected to the jet. The
radio/X-ray ratio in the secondary hotspot region is a factor $\ga 50$
greater than in knot B. There is no optical detection of the secondary
\citep{pk97}, nor is it detected in the archival {\it
Spitzer} data.
\label{nx2}
We extracted spectra for the jet, for the counterpart to knot B (NX1)
as a whole, for its two subcomponents (denoted NX1E and NX1W) and for
the possibly unrelated source NX2, taking background from a nearby
blank-sky region. The overall spectrum of NX1 is not particularly well
fitted with a single power-law model ($\chi^2 = 15.9/8$; $\Gamma =
1.95 \pm 0.15$). The E component of the X-ray source is well fitted
with a power-law model ($\chi^2 = 1.8/3$, $\Gamma = 2.23 \pm 0.23$)
but the W component is not ($\chi^2 = 10.5/3$, $\Gamma = 1.7 \pm
0.2$), the poor fit being the result of one high bin at soft energies.
Neither component is acceptably fitted with a thermal model. Within
the limited statistics, it seems likely that the two components of
knot B have significantly different X-ray spectra. The total 1-keV
unabsorbed flux density of knot B in our extraction is $4.2 \pm 0.3$
nJy, with the two sub-components being roughly equal in flux. The few
counts in the jet are well fitted ($\chi^2 = 0.4/1$) with a power-law
model with $\Gamma = 1.4 \pm 0.4$, and a flux density $1 \pm 0.2$ nJy.
Finally, NX2 is well fitted ($\chi^2 = 1.5/2$) with a power-law model
with $\Gamma = 2.13 \pm 0.24$ --- comparable to the spectral indices
of NX1 or its components --- and has a 1-keV flux density of $1.9 \pm
0.2$ nJy.

Fig.\ \ref{390.3hss} shows the X-ray emission from the southern
hotspot. The brighest compact feature, denoted SX1 in the figure, is
not coincident with any named radio feature, but appears at the N end
of the bright `rim' of the hotspot discussed by \citet{lp95}. The
nearest discrete compact feature, $1\farcs0 \pm 0\farcs2$ further
around the rim, is a weak radio knot (which we denote G$'$) visible in
the high-resolution radio map of \citet{lp95}. This feature
may be considered the primary hotspot or may just be a jet knot.
Otherwise there is no particular association between the detected
X-ray emission and the radio: none of the bright knots G, E, D clearly
has a compact associated X-ray source, though all have X-ray emission
coincident with them, and the X-rays from the `tail' region bear no
relation to the radio structure seen here. Archival {\it Spitzer}
images show that the S hotspot has a 24-$\mu$m infrared counterpart,
which appears to be peaked where the radio is brightest (i.e. in
hotspots E and D), but the low resolution of {\it Spitzer} at this
wavelength prevents us from examining this in detail. Taking a single
spectrum of all the detectable X-ray structure in Fig.\
\ref{390.3hss}, we find it to be poorly fitted ($\chi^2 = 11/5$) with
a single power law with $\Gamma = 1.4 \pm 0.2$ and a total 1-keV flux
density of $2.7 \pm 0.3$ nJy. The poor fit probably indicates that
more than one spectral component is present, but there are too few
counts to try to separate these spatially. Using the parameters of
this fit, the compact component SX1 would have a flux density $\sim
0.4$ nJy.

\subsection{3C\,403}

The hotspots of 3C\,403 were discussed extensively by \citet{khwm05},
and so we do not present images or spectral fits here. To summarize
the results that are most important for this paper, the hotspots F1
and F6 in the E lobe have clear X-ray detections, with measured photon
indices $\sim 1.7$. F6 may be a jet knot, consistent with the
detection of jet-related X-ray emission in the same lobe, and if so F1
is the primary hotspot. The peaks of the X-ray emission from F1 and F6
are clearly coincident with the peaks in the radio, unlike several of
the other hotspots considered in this paper, but both have X-ray
extensions back towards the nucleus that are not present (or at least
not nearly as prominent) in the radio. There is no detection of the
radio-bright secondary hotspots F2 and F3 and there is no diffuse
X-ray emission from the hotspot region.

\section{Discussion}

\subsection{Offsets and emission mechanisms}

The observations of 3C\,227, 327 and 390.3 almost all show differences
between the structure observed in the radio and that observed in the
X-ray. To summarize, the only `simple' hotspots seen in these three
objects are the weak compact components in the E hotspot of 3C\,227 and
(probably) the S hotspot of 3C\,390.3, SX1. 3C\,227 shows a striking 1.5-kpc
offset between the radio emission from the primary W hotspot and its
X-ray counterpart, in the sense that the X-ray emission is nearer the
nucleus, while having similar structure in the radio and X-ray. The
well-known counterpart to 3C\,390.3's hotspot B turns out to have two
components, neither of which is coincident with the radio or optical
detection, and which are displaced from it in a direction
perpendicular to the jet axis. Neither of the two features most
obviously related to the E hotspots of 3C\,327 is coincident with the
peak of the radio: one is 4 kpc along the jet axis from the primary
hotspot, the other at least 1 kpc away from the secondary along the
primary-secondary axis.

What physical processes can account for these diverse behaviors? One
possibility is of course that the compact components are nothing to do
with the radio sources. This is our preferred explanation for the
compact feature NX2 in 3C\,390.3 N (\S\ref{nx2}): it may also explain
one or, at a stretch, both of the features we see in 3C\,327 E, where
we have no deep optical information to constrain whether a background
object could be responsible. It seems very unlikely that it could
account for the observations of 3C\,390.3 NX1 or 3C\,227, where the
X-ray features are clearly resolved. We therefore consider the
possible emission processes that could be responsible for producing
the observed X-rays from the sources themselves. As the available
X-ray data suggest power-law spectra and thus non-thermal emission, we
focus on inverse-Compton and synchrotron processes.

As discussed by \citet{hbch02} in the context of the more distant
double-hotspot source 3C\,351 (where inverse-Compton emission almost
certainly plays some role in the complex X-ray structures seen),
synchrotron self-Compton emission (SSC) or inverse-Compton scattering
of the CMB (CMBIC) cannot produce offsets between the radio and X-ray
unless there is very strong spatial variation in the positions of the
low-energy electrons and/or important beaming effects. We first
consider a model in which there are strong point-to-point variations
in the number density of low-energy electrons. To produce offset X-ray
emission via inverse-Compton processes, we require a large population
of low-energy ($\gamma \la 10^3$) electrons at the location of the
X-ray emission, while either the electron spectrum or the magnetic
field strength must be tuned so as to avoid significant emission from
this population of electrons at radio frequencies. We used the code of
\citet{hbch02} to calculate the expected inverse-Compton emission from
components matching P1, P2 and the X-ray source in 3C\,227, using an
upper limit on the radio flux density of the X-ray component of 2 mJy
to normalize the radio spectrum and using the observed sizes in the
radio and X-ray to choose component sizes. We modeled the three
components for simplicity as uniformly filled ellipsoids in the plane
of the sky. We found that for standard broken power-law spectra for
the three components we require the X-ray source to have a departure
from equipartition of a factor $\ga 60$ in magnetic field strength,
and an electron energy density that is $\ga 10^3$ times greater than
that in P2, in order to produce the observed X-rays by inverse-Compton
processes (where the upper limits come from the fact that we have no
unambiguous detection of a radio counterpart of the X-ray source). In
this situation we find that the dominant photon field is the CMB, and
so the conclusions are robust against uncertainties in the geometry,
which affect only the number density of synchrotron photons from P1/2.
Such an electron distribution seems highly unlikely simply on
energetic grounds, since it requires essentially all the energy in the
hotspot to be concentrated in an offset, radio-invisible region. A
similarly implausible electron distribution would be required for
3C\,327 and 3C\,390.3. In addition, we show in Table \ref{hsic} that
the parameters of all detected and non-detected hotspots predict
inverse-Compton flux densities (derived using the code of
\citealt{hbw98}) for equipartition magnetic fields that are much less
than the observed values or limits, as was shown previously in some
cases by \citet{hhwb04}. We therefore rule out simple inverse-Compton
models in what follows.

\begin{deluxetable}{lllrrrrrrl}
\tablewidth{21cm}
\tabletypesize{\footnotesize}
\rotate
\tablecaption{Inverse-Compton flux densities and radio/X-ray ratios
  for the hotspots}
\label{hsic}
\tablehead{\colhead{Source}&\colhead{Lobe}&\colhead{Radio}&\colhead{Size}&
\colhead{Radio frequency}&\colhead{Radio flux}&\colhead{1-keV
  flux}&\colhead{Predicted IC flux}&\colhead{X-ray/radio}&\colhead{Notes}\\
&&\colhead{name}&\colhead{(arcsec)}&\colhead{(GHz)}&\colhead{(mJy)}&
  \colhead{(nJy)}&\colhead{(nJy)}&\colhead{($\times 10^{6}$)}\\
}
\startdata
3C\,227&W&P1&$1.5 \times 0.5$&8.35&16&$1.5 \pm 0.2$&0.0010&0.09&Offset X-ray\\
&&P2&$3.5 \times 0.5$&8.35&25&$1.5 \pm 0.2$&0.0019&0.06&Offset X-ray\\
&&P3&$3.7 \times 0.6$&8.35&13&$0.3 \pm 0.1$&0.0014&0.02&Offset X-ray\\
&&P4&$1.4 \times 0.25$&8.35&1.2&$0.4 \pm 0.1$&$1.0 \times 10^{-4}$&0.3&Jet knot?\\
&E&F1&$1.2 \times 0.25$&8.35&5.3&$0.3 \pm 0.1$&$2.7 \times 10^{-4}$&0.06\\[4pt]
3C\,327&E&S1&$1.9 \times 0.25$&8.35&15&$0.27 \pm 0.07$&$6.4 \times 10^{-4}$&0.02&Offset X-ray\\
&&S2&0.32&8.35&3.2&$0.27 \pm 0.07$&$1.2 \times 10^{-4}$&0.09&Offset X-ray\\[4pt]
3C\,390.3&N&B&$1.3 \times 0.5$&4.99&66&$4.2 \pm 0.3$&0.003&0.08&2 sub-components\\
&&F&3.7&4.99&190&$<0.2$&0.009&$<0.001$&Upper limit\\
&S&G$'$&0.5&4.99&20&$0.4 \pm 0.1$&$6 \times 10^{-4}$&0.02&Offset X-ray\\
&S&G&0.5&4.99&36&$<0.1$&0.0013&$<0.004$&Upper limit\\
&S&E&0.7&4.99&106&$<0.1$&0.0056&$<0.001$&Upper limit\\
&S&D&1.2&4.99&206&$<0.2$&0.011&$<0.001$&Upper limit\\[4pt]
3C\,403&E&F1&0.275&8.35&16&$0.9 \pm 0.2$&$5 \times 10^{-4}$&0.04\\
&&F1b&0.256&8.35&7.7&$0.5\pm 0.1$&$2 \times 10^{-4}$&0.06&Jet knot?\\
&&F2&$1.8 \times 0.25$&8.35&27&$<0.13$&0.0011&$<0.004$&Upper limit\\
&&F6&0.272&8.35&27&$2.3 \pm 0.2$&0.0014&0.06&Jet knot\\
\enddata
\tablecomments{\scriptsize Columns 1, 2 and 3 are as in Table \ref{hstable}.
  Column 4 gives the size used in modeling, derived from fits to the
  high-resolution radio data. The angular sizes used are the radii of
  homogenous sphere models fitted to the data, as described by
  \citet{hhwb04}, except where two numbers are quoted, in which case
  they are the length and radius of a cylinder. Columns 5 and 6 give
  the frequency and flux of the radio data used to normalize the
  inverse-Compton models. Column 7 is the 1-keV X-ray flux density of the X-ray
  counterpart, where present, taken from or derived as in Table
  \ref{hstable}. Column 8 gives the 1-keV inverse-Compton prediction (SSC+CMBIC). Column 9 gives the ratio of the X-ray and
  radio flux densities, corrected to 8.35 GHz assuming $\alpha = 0.5$. Column 10 gives any comments on the
  relationship between the X-ray flux quoted and the radio hotspot for
  which the inverse-Compton calculation was made. Data for 3C\,403 are
taken from \citet{khwm05}.}
\end{deluxetable}

\citet{gk03} (hereafter GK03) proposed to explain the X-ray properties
of hotspots in objects aligned close to the line of sight using a
model involving emission from the decelerating relativistic jet. At
the time they were writing, there was an apparent correlation between
observations of hotspots with X-ray emission too bright to be SSC in
origin and the jet side of broad-line objects, such as broad-line
radio galaxies and radio-loud quasars, that are expected to lie at
small angles to the line of sight in unified models. Since then,
further observations have established that these non-SSC hotspots can
occur in objects in all orientations \citep{hhwb04}. However, the GK03
model is still interesting because it predicts spatial offsets between
the peak of the X-ray and that of the radio. They proposed that the
jet decelerates on kpc scales and that X-ray emission from the
fast-moving component can be produced by inverse-Compton scattering of
the synchrotron photons from the slower, downstream component. This
external inverse-Compton process is strongly directional and can only
be seen if the jet is aligned close to the line of sight (and, of
course, is approaching rather than receding). The GK03 model should
not be confused with the process of inverse-Compton scattering of the
CMB that has been proposed to explain the X-ray jets in core-dominated
quasars \citep[e.g.,][]{tmsu00}: in the GK03 model the photon energy
density is dominated by synchrotron photons from the upstream hotspot,
and as a consequence the beaming factors required (for given jet
properties) are less extreme and the range of plausible angles to the
line of sight can be greater. In the beamed CMB models the X-ray
emission should have no particular relationship to the radio hotspot,
and the angle to the line of sight of the jet is required to be small:
we therefore do not consider these models further.

The GK03 model might be applied to the offset between the primary
hotspot of 3C\,227, P1/2, and its X-ray counterpart. 3C\,227 is a
broad-line object, so that the lobes make a relatively small angle to
the line of sight ($\la 45^\circ$) and, if we take P4 to be a jet
knot, the only evidence that we have suggests that the W lobe is
likely to contain the jet pointing towards us, though the
comparatively weak radio core, the non-detection of any other
components of the jet, and the presence of intrinsic absorption in the
X-ray spectrum (\S\ref{227-core}) suggest that it is not at a very
small angle to the line of sight (compared to, say, 3C\,390.3, with
its bright radio core, unabsorbed nuclear X-ray emission and
well-detected kpc-scale jet). The flat X-ray spectrum of the X-ray
counterpart of P1/2 is consistent with an inverse-Compton model. If we
suppose that the jet decelerates on scales of a few kpc (set by the
observed projected size of the offset between radio and X-ray) from
relativistic to sub-relativistic speeds, is it possible that the GK03
model could explain the observed offset in 3C\,227? In principle the
answer is `yes' since the GK03 model can explain {\it any} observation
with a suitable choice of the (observationally unconstrained)
parameters of the position-dependent bulk flow speed and electron
energy spectrum of the jet. The jet would have to be quite wide at the
position of P1/2 to produce the observed X-ray morphology, but not
impossibly so: the distribution of emitting particles and jet
velocities would also require some fine tuning in order to produce an
offset X-ray peak rather than some more jet-like X-ray structure.
However, we consider an explanation in terms of the GK03 model to be
hard to sustain when all the observations of our sample are considered
together. The secondary hotspot of 3C\,227 shows a similar offset, yet
the bulk flow clearly cannot decelerate to sub-relativistic speeds
twice --- and the direction of any flow between the primary and
secondary is not the same as the direction between P4 and P1/2. A
series of coincidences is required to explain the similarity between
the primary and secondary hotspots in 3C\,227, therefore. The model
clearly cannot explain the offsets between radio and X-ray peaks in
3C\,390.3 (not in the direction of the jet) or the offsets and
variations in radio/X-ray ratio seen in 3C\,327 and 3C\,403 (too large
an angle to the line of sight). We conclude that the GK03 model, while
not ruled out by the data for the primary hotspot of 3C\,227, is of
little use in providing a general explanation of the problems posed by
our observations.

Since inverse-Compton explanations seem difficult to accept, we next
consider synchrotron emission. Synchrotron explanations also require
point-to-point electron spectrum (or possibly magnetic field strength)
variations to account for offsets between the radio and X-ray
emission. However, the magnitude of the variation is comparatively
very small: the high-energy tail of the electron population
responsible for the X-ray emission is energetically, and a fortiori
numerically, a negligible fraction of the total, whereas for the
inverse-Compton process a doubling of the emissivity by adjusting the
electron spectrum requires a doubling of the number density of
low-energy electrons, and so effectively a doubling of the energy
density of the system. Energetically, therefore, it is not difficult
to produce what we see via synchrotron radiation. Since the electron
energy loss timescale is likely to be very short (for field strengths
close to the equipartition value) what is required to produce the
observed X-ray structures is some process that can accelerate
particles wherever we observe X-ray emission. As we discussed in
\S\ref{intro}, there is strong evidence from radio and optical data
that hotspots are sites of particle acceleration, and their spectra
are consistent with models involving a single shock followed by
downstream losses. In some cases the X-ray emission from hotspots lies
on an extrapolation of these models \citep{khwm05}. But observations
of {\it diffuse} X-ray emission, often poorly matched to the observed
radio structures (e.g. in Pictor A, \citealt{hc05}; 3C\,33,
\citealt{kbhe07}; 3C\,390.3 S, this paper) make it hard to sustain a
model in which the particle acceleration at the hotspots is {\it only}
occurring at jet termination shocks. Similar conclusions have been
reached by other authors based on optical data \citep[e.g.,][]{rm87,pbm02}.

Observations of compact but offset X-ray emission, as in 3C\,227,
present a different problem. Here it seems possible that there is a
discrete acceleration region that is related to the jet termination
shock, but, if so, the shock is not where we would have inferred it to
be from radio observations. For 3C\,227 we could imagine a picture in
which the X-ray emission in both primary and secondary hotspots tells
us where the shock is now, while the radio traces material that has
passed through this shock region, expanded and decelerated. This would
imply that the X-ray emission should have a radio/optical counterpart,
but for a flat synchrotron spectrum ($\alpha \sim 0.5$) extending
between radio and X-ray the emission at other wavebands could be
undetectably faint, at the 10 $\mu$Jy level in the radio (i.e.,
substantially below the upper limit of $\sim 2$ mJy on the flux
density of this component from radio maps). Such a flat
synchrotron spectrum extending all the way to the X-ray has never been
observed (precisely because of the difficulty of detecting the radio
counterparts) but might be expected in models of shock acceleration.
The bulk of the optical emission in 3C\,227 P1/2 seems to lie in
between the X-ray and radio peaks, which is qualitatively consistent
with this picture. The questions to be asked are then why other
hotspots seen with similar spatial resolution, such as those in
3C\,403, do not show the same radio/X-ray offsets; why the offsets in
3C\,327, if they have the same origin, are so much larger; and what
the origin is of the compact structure transverse to the jet direction
in the hotspot of 3C\,390.3.

Recent numerical simulations suggest that the picture of particle
acceleration in the lobes of FRII sources may be less simple than in
the traditional model of acceleration at strong shocks in one or more
hotspots. \citet*{tjr01} carried out three-dimensional
MHD simulations that modeled the transport of relativistic electrons
and of particle acceleration at shocks. They found that the
interaction of the jet and the backflowing plasma at the head of the
jet produced what they called a `shock-web complex', ``a region of
shocks of varying strengths and sizes spread throughout the source''.
Even when there was a simple terminal shock, not all the jet material
necessarily passed through it, and the terminal shock was not always
the strongest shock in the system. While it is not clear that their
simulations are perfectly matched to real radio sources, they are
capable of producing simulated synchrotron images that show apparent
clear discrete multiple hotspots \citep{tjrp02} and in these
cases the particle acceleration is not necessarily well matched to the
locations of the hotspots: hotspot locations in their model can have
more to do with magnetic field amplification than with particle
acceleration. The notion of a `shock-web complex' at the head of the
jet could help to explain the diffuse X-ray emission now seen in the
radio-bright but non-compact source head regions of a number of
objects, as discussed above, while the idea that the particle
acceleration region may not always be co-spatial with the observed
radio hotspot might help to explain observed offsets. It should be
possible to carry out numerical simulations that allow synthetic maps
of the location of high-energy synchrotron-emitting particles to be
generated for qualitative comparison with the range of structures seen
in X-ray observations.

We can also compare the X-ray observations of hotspots with
observations of systems in which the X-ray emission is almost
certain to be synchrotron in origin, the FRI jets (\S\ref{intro}). In
the nearest FRI jet, Centaurus A \citep{hwkf03} there is
direct dynamical evidence for shock-acceleration of particles, as we
believe is going on in FRII hotspots. There are also offsets, albeit
on scales of only tens of pc rather than kpc, between the peak of the
X-ray emission and the brightest radio emission, in the sense that the
radio emission peaks downstream of the X-ray. And there is diffuse
X-ray emission, not associated with any compact radio source or
dynamical feature of the jet, which in some cases has a diffuse optical
counterpart \citep{hkw06}. Other FRI jets show
similar features. At present we do not understand the nature of the
radio/X-ray peak offsets in FRI jets or the distributed particle acceleration
process responsible for the diffuse X-ray emission, but the
qualitative similarity between the Cen A jet and a jet termination
region like 3C\,390.3 S or 3C\,33 S means that we might hope to gain
some insight into one problem by studying the other. In both cases the
observational requirement is sensitive, multi-frequency observations
that allow us to construct a detailed map of the synchrotron SED as a
function of position. 

\subsection{The nature of multiple hotspots}

Our two new targets, 3C\,227 and 3C\,327, provide at least one clear
example (3C\,227 W), and possibly two, of an object where the primary
and secondary hotspot are both detected in the X-ray, setting aside
the problem of offsets between the components. 3C\,33 N \citep{kbhe07}
is another example of a source with multiple X-ray hotspots. 3C\,390.3
N, on the other hand, behaves more similarly to 3C\,403 E: the bright
secondary hotspot is not an X-ray (or optical) synchrotron source even
though there is an apparently clear connection between the primary and
secondary hotspot indicative of continuing energy supply (but cf.\ the
discussion of this point in \citealt{lp95}). Taking SX1 and its radio
counterpart G$'$ to be the primary hotspot of the southern hotspot
complex, a similar statement can be made for this system too. In both
these cases, the upper limit on the X-ray to radio flux ratio in the
non-detected hotspots, which are generally brighter in the radio, is
1--2 orders of magnitude below the measured value for the primary
hotspots (Table \ref{hsic}). If we assume, as discussed in the
previous section, that the X-ray emission mechanism is synchrotron,
then this tells us that secondary hotspots can be different: some, at
least, are able to accelerate particles to the highest observable
energies, but others are at least an order of magnitude less efficient
than the primaries in producing X-ray emission for a given radio
emissivity. This conclusion would be stronger if the nature of the
X-ray emission in the secondary hotspots were more obvious.

If some secondary hotspots can accelerate particles to high energies
and some do not, what is the difference between them? Relic hotspots
left behind by a jet that has moved (`dentist's-drill' model) would
certainly not be expected to have high-energy particle acceleration.
But in our observations one secondary that apparently is connected to
the jet (3C\,390.3 N) does not have high-energy particle acceleration,
while one that has no apparent connection in sensitive radio
observations (3C\,227 W) does. Radio morphology is therefore not a
good guide to a hotspot's ability to accelerate particles, or to its
relationship to the energy supply. Nor is the radio brightness of the hotspot.

One trend that is apparent in the data is that a hotspot is more
likely to be an X-ray emitter (and therefore a privileged site for
high-energy particle acceleration?) if it is compact. The secondary
hotspots in 3C\,227 and 3C\,327 are similar in size to the primaries.
Those in 3C\,390.3 and 3C\,403 are several times larger. `Compact'
here appears to mean less than a few kpc in size. However, though this
may be a necessary condition, it is not a sufficient one, as the
non-detection of relatively compact hotspots in e.g. 3C\,390.3 S
shows. Secondary hotspot compactness could thus be an indicator of
relatively well-collimated continued outflow from the primary hotspot
to the secondary (or, in the case of the \citealt{cgs91} model, of a
well-collimated disconnected jet); this makes sense, since (for a
given luminosity) a more compact hotspot is more overpressured with
respect to the lobe material and will have a shorter timescale for
disappearance via adiabatic expansion in the absence of the energy
supply. But the lack of a one-to-one correlation reinforces what we
already know from observations of single hotspots: the ability of even
low-luminosity, low-$B$-field hotspots, even when clearly connected to
the energy supply, to produce X-ray emission is very variable and must
depend on details of the microphysics that are not yet accessible to
us.

\section{Conclusions}

We have looked with very high spatial resolution at the hotspot X-ray
emission from a small sample of radio galaxies that show multiple
radio hotspots. As in earlier work, we argue that the X-ray
emission from the hotspots comes predominantly from the synchrotron
process, and so traces high-energy particle acceleration. To our
knowledge this paper represents the first attempt to use synchrotron
emission to probe the particle acceleration properties in a sample of
FRII sources, though several individual objects have previously been
studied in detail.

Our principal results can be summarized as follows:

\begin{itemize}
\item The cores and lobes of the two new sources in our sample have
  X-ray properties that are entirely consistent with expectations and
  with the sources' places in unified models. There is evidence for
  intrinsic absorption in the spectrum of the BLRG 3C\,227.

\item All the target sources exhibit structure in the X-ray images of
  their hotspots that would not have been predicted in a simple model
  in which particle acceleration occurs only at the jet termination as
  traced by the bright radio hotspot. This structure ranges from
  small-scale offsets in the radio and X-ray peaks (e.g. in 3C\,227 W
  or 3C\,390.3 N) through diffuse X-ray emission that is not well
  correlated with compact radio structure (e.g. 3C\,390.3 S: see also
  3C\,33 S, \citealt{kbhe07}; Pictor A E, \citealt{hc05}) to
  point-like sources in the jet termination region that bear little
  obvious relationship to the current radio hotspots (3C\,327 E). If
  most or all of these structures can be taken to indicate the
  location of particle acceleration in these sources, then our
  observations support models in which the particle acceleration
  history in FRIIs can be complicated, non-localized, and not well
  traced by radio observations.

\item Our observations were obtained to investigate the nature of
  multiple hotspots, and we have found some evidence that some
  secondary hotspots are indeed associated with acceleration of
  particles to the highest observable energies, while others (as we
  had found previously) are not. This implies that at least some
  secondary hotspots have ongoing access to a supply of energy. All
  X-ray-synchrotron emitting hotspots appear to be compact, but not
  all compact hotspots are detected in X-rays. We cannot at present
  say whether this is because some of these compact hotspots are true
  relics, disconnected from the energy supply, or whether they are
  X-ray faint for some other reason related to the microphysics of
  their particle acceleration. Sensitive multi-wavelength observations
  in radio and optical will be required to make further progress.
\end{itemize}

\acknowledgements

We are very grateful to Karl-Heinz Mack for providing us with optical
images of the hotspots of 3C\,227 prior to publication and for helpful
discussion of the radio-optical alignment. We thank an anonymous
referee for constructive comments that helped us to improve the paper.
We also gratefully acknowledge financial support for this work from
the Royal Society (research fellowship for MJH) and NASA (grant
GO6-7094X to RPK).

The National Radio Astronomy Observatory is a facility of
the National Science Foundation operated under cooperative agreement
by Associated Universities, Inc.

\end{document}